\documentclass[aps,prl,twocolumn,superscriptaddress]{revtex4-2}

\usepackage{amsmath,amssymb,bm}
\usepackage[T1]{fontenc}
\usepackage{graphicx}
\usepackage{physics}
\usepackage{braket}
\usepackage{xcolor}
\begin{document}

\title{
Breakdown of Fluctuational Electrodynamics in the Extreme Near Field
}

\author{Philippe Ben-Abdallah}
\email{pba@institutoptique.fr}
\affiliation{
Laboratoire Charles Fabry, UMR 8501, Institut d'Optique, CNRS,
Universit\'e Paris-Saclay,
2 Avenue Augustin Fresnel, 91127 Palaiseau Cedex, France
}

\date{\today}

\begin{abstract}
Fluctuational electrodynamics relies on the assumption that thermal
fluctuations in distinct bodies are statistically independent.
It is shown that this approximation ceases to be valid in the extreme
near-field regime, where hybridization of surface phonon-polaritons
across nanometric vacuum gaps generates finite fluctuating-current
cross correlations between opposite interfaces.
Using a microscopic coupled-oscillator model combined with a
Green-tensor formulation of the Poynting vector, the resulting
correlation-induced correction to radiative heat transfer is derived.
For polar materials, these correlations become significant when the
hybridization energy approaches the intrinsic damping rate and can
substantially modify conventional fluctuational-electrodynamics
predictions at subnanometric separations.
\end{abstract}

\maketitle

Radiative heat transfer at subwavelength distances is governed by
thermally fluctuating electromagnetic fields generated inside condensed
materials. In the near-field regime, evanescent electromagnetic modes
dominate the energy exchange and can produce heat flux exceeding the
blackbody limit~\cite{Planck} by several orders of magnitude. These
enhancements of radiative heat transfer originate from the tunneling of
evanescent electromagnetic modes across nanoscale vacuum
gaps~\cite{JoulainSurfSciRep05,VolokitinRMP07,Cuevas,RMP}. Such phenomena
are commonly described within the framework of fluctuational
electrodynamics (FE) theory, originally developed by
Rytov~\cite{Rytov}.
A central assumption of FE is that fluctuating thermal sources located
inside distinct bodies remain statistically independent. Within this
framework, near-field heat transfer is interpreted as a purely
electromagnetic process mediated by the exchange of evanescent photons
between independent thermal emitters.
At nanometric and subnanometric separations, however, the evanescent
fields of opposite interfaces strongly overlap, and their surface
excitations hybridize into collective symmetric and antisymmetric modes.
Although this deterministic hybridization is already contained in the
Green tensor of conventional FE, it raises a distinct question: can the
effective fluctuating surface degrees of freedom still be described as
statistically independent sources?
This question differs from previously discussed limitations of local FE
at microscopic separations, such as nonlocal response or electron and
phonon tunneling~\cite{GomezViloria,Guo2023,GomezViloriaElectron}, which modify the electromagnetic susceptibility or
open additional transport channels.

In this Letter, instead, we investigate the
cross-correlations of the interacting surface fluctuations.
Using a coupled-mode description driven by independent Langevin forces,
we show that the interaction generates a finite off-diagonal covariance
between the effective surface-mode currents. This covariance is distinct
from correlations between fields or induced currents already contained
in conventional FE. Combined with a Green-tensor formulation of the
Poynting vector, it yields a correlation-induced contribution to the
radiative heat flux, which becomes significant when the
hybridization-induced mode splitting approaches the intrinsic linewidth.

To describe this regime, we model the surface optical phonons by two coupled oscillators characterized by generalized displacement coordinates $X_1$ and $X_2$,
\begin{equation}
D_1(\omega)X_1+KX_2=\xi_1,
\label{eq:eq1}
\end{equation}
\begin{equation}
KX_1+D_2(\omega)X_2=\xi_2,
\label{eq:eq2}
\end{equation}
where
\begin{equation}
D_i(\omega)
=
M_i
\left(
\Omega_i^2-\omega^2-i\gamma_i\omega
\right)
\label{eq:Di}
\end{equation}
is the phonon dynamical response function. Here $M_i$ is the effective mass associated with the surface optical phonon mode in body $i$, $\Omega_i$ its resonance frequency, and $\gamma_i$ the damping rate describing phonon dissipation inside the material.
The deterministic and stochastic parts of this construction must be
distinguished. The matrix formed by $D_1$, $D_2$, and $K$ determines the
normal-mode spectrum, whereas the Langevin correlators specify their
thermal excitation. In particular, $K$ is determined independently from
the Maxwell surface-mode dispersion, while
$\langle \xi_1\xi_2^*\rangle=0$.

The coupling coefficient $K$ originates from the overlap of the evanescent surface phonon-polariton fields localized near the two interfaces. In the weak-coupling regime, the interaction decays exponentially with separation distance because of the evanescent character of the surface fields. In the extreme near-field regime considered here, however, where $k_\parallel d\lesssim1$, the coupling strength must instead be extracted from the full hybridized surface phonon-polariton dispersion relation discussed below. The Langevin forces satisfy~\cite{Callen,Kubo}
\begin{equation}
\left\langle
\xi_i(\omega)\xi_j^*(\omega')
\right\rangle
=
2\pi
\delta(\omega-\omega')
S_i(\omega)\delta_{ij},
\end{equation}
with
\begin{equation}
S_i(\omega)
=
2M_i\gamma_i\hbar\omega
\coth
\left(
\frac{\hbar\omega}{2k_BT_i}
\right).
\end{equation}
Solving Eqs.~(\ref{eq:eq1}) and (\ref{eq:eq2}) gives
\begin{equation}
X_i
=
\frac{
D_j\xi_i-K\xi_j
}{
D_iD_j-K^2
},
\qquad
(i\neq j).
\end{equation}
The displacement cross correlation therefore reads
\begin{equation}
\left\langle
X_iX_j^*
\right\rangle
=
\frac{
\left\langle
(D_j\xi_i-K\xi_j)
(D_i^*\xi_j^*-K\xi_i^*)
\right\rangle
}{
|D_iD_j-K^2|^2
}.
\end{equation}
Using the statistical independence of the Langevin forces,
\begin{equation}
\left\langle
\xi_i\xi_j^*
\right\rangle
=
0,
\qquad
(i\neq j),
\end{equation}
one obtains
\begin{equation}
\left\langle
X_iX_j^*
\right\rangle
=
-
K
\frac{
D_jS_i+D_i^*S_j
}{
|D_iD_j-K^2|^2
},
\qquad
(i\neq j).
\label{eq:cross}
\end{equation}
The $K$-dependent terms describe fluctuations transferred between the
two coordinates by deterministic coupling; all correlations share the
same denominator because they probe the same hybridized normal-mode
poles.

The fluctuating dipole moments associated with the surface optical phonons are written as
\begin{equation}
\mathbf p_i(\omega)
=
\mathbf Q_iX_i(\omega),
\end{equation}
where $\mathbf Q_i$ denotes the effective Born charge vector associated with the surface mode.
The corresponding fluctuating current is obtained from the time derivative of the dipole moment,
\begin{equation}
\mathbf J_i(\omega)
=
-i\omega
\mathbf Q_iX_i(\omega).
\label{eq:JX}
\end{equation}
The cross-correlation of fluctuating currents is therefore directly related to the displacement cross-correlation,
\begin{equation}
\mathbf C_{12}^{J}
\equiv
\left\langle
\mathbf J_1
\otimes
\mathbf J_2^\dagger
\right\rangle
=
\omega^2
\left\langle
X_1X_2^*
\right\rangle
\mathbf Q_1
\otimes
\mathbf Q_2^\dagger.
\label{eq:JcorrXcorr}
\end{equation}
Substituting Eq.~(\ref{eq:cross}) yields
\begin{equation}
\mathbf C_{12}^{J}
=
-\omega^2
K
\frac{
D_2S_1+D_1^*S_2
}{
|D_1D_2-K^2|^2
}
\mathbf Q_1
\otimes
\mathbf Q_2^\dagger.
\label{eq:currentcross}
\end{equation}

Importantly, $\mathbf C_{12}^{J}$ is interpreted here as an
off-diagonal covariance of the effective fluctuating surface-mode
sources, rather than a correlation between induced currents generated
by multiple scattering, which is already contained in conventional FE.
Writing
\begin{equation}
\mathbb C_{\rm src}
=
\mathbb C_{\rm diag}
+
\begin{pmatrix}
0 & \mathbf C_{12}^{J}\\
\mathbf C_{21}^{J} & 0
\end{pmatrix},
\label{eq:Cblock}
\end{equation}
conventional FE corresponds to the diagonal sector dressed by the full
Green tensor, whereas the present contribution originates from the
off-diagonal sector. The two therefore represent distinct statistical
contractions despite sharing the same hybridized electromagnetic
structure.

\begin{figure}
\centering
\includegraphics[angle=0,scale=0.4]{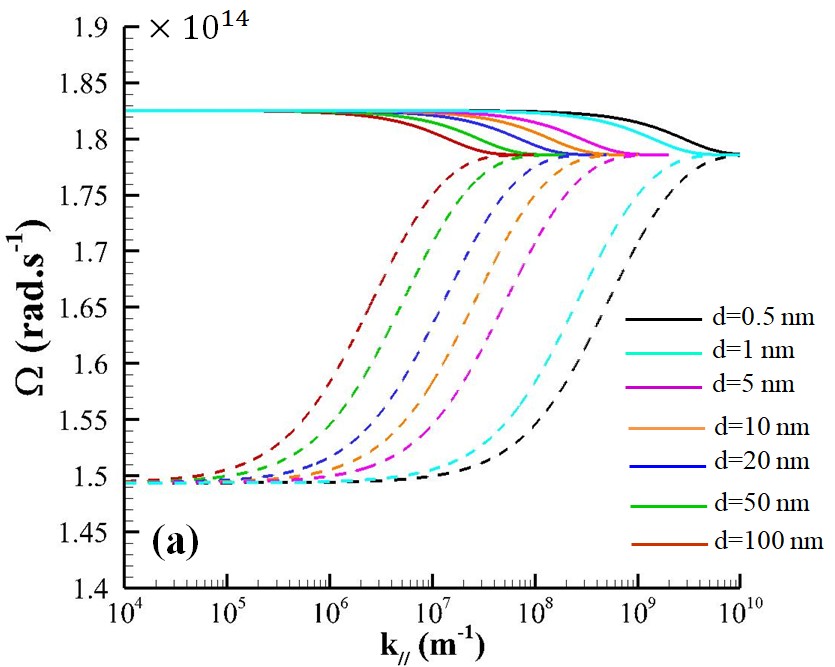}
\includegraphics[angle=0,scale=0.4]{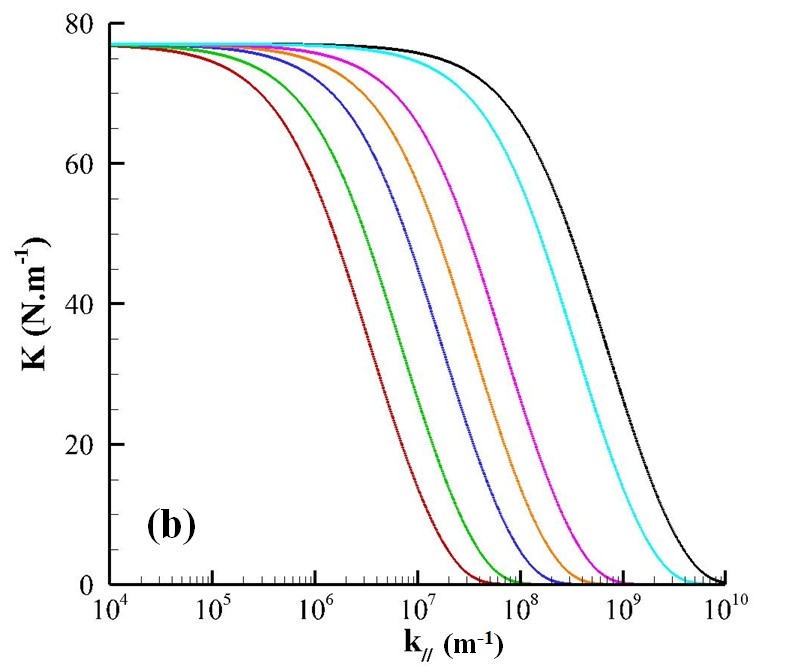}
\caption{(a) Resonance frequency of hybridized surface phonon-polaritons between two SiC half-spaces separated by a vacuum gap $d$. In full line $\Omega_+$ and in dashed line $\Omega_-$ (b) Interfacial coupling coefficient $K(k_\parallel,d)$ between the two surface phonon-polariton oscillators localized the opposite interfaces for different separation distances. 
}
\end{figure}

The spectral heat flux density is obtained from the ensemble-averaged
Poynting vector. For planar geometries,
\begin{equation}
\varphi(\omega)
=
\left\langle
\Pi_z(\omega)
\right\rangle,
\end{equation}
where the electric and magnetic fields generated by the fluctuating
currents are expressed through the electromagnetic Green tensors.
Substituting these fields into the Poynting vector yields a bilinear
form involving the current-current correlation tensor
\(
\langle \mathbf J\otimes\mathbf J^\dagger\rangle
\).
Decomposing the fluctuating currents as
\begin{equation}
\mathbf J
=
\mathbf J_1+\mathbf J_2,
\end{equation}
one obtains both autocorrelation and cross-correlation contributions.
In conventional FE, the off-diagonal
correlations vanish,
\begin{equation}
\mathbf C_{12}^{J}
=
\mathbf C_{21}^{J}
=
0,
\end{equation}
so that heat transfer originates solely from independent fluctuating
sources. For planar geometries, the spectral heat flux takes the
Landauer form~\cite{PoldervH},
\begin{equation}
\begin{aligned}
\varphi_{\mathrm{FE}}(\omega)
=&\,
\frac{1}{2\pi}
\left[
\Theta(\omega,T_1)
-
\Theta(\omega,T_2)
\right]
\\
&\times
\sum_{\alpha=s,p}
\int_0^\infty
k_\parallel\,dk_\parallel\,
\mathcal T_{\alpha}(\omega,k_\parallel,d).
\end{aligned}
\label{eq:phiFE_spectral}
\end{equation}
where \(\mathcal T_\alpha\) denotes the usual transmission coefficient.
For evanescent modes,
\begin{equation}
\mathcal T_{\alpha}
=
\frac{
4\,\operatorname{Im}(r_{1\alpha})
\operatorname{Im}(r_{2\alpha})
e^{-2k_\parallel d}
}{
\left|
1-r_{1\alpha}r_{2\alpha}e^{-2k_\parallel d}
\right|^2
},
\label{eq:transmission}
\end{equation}
with \(r_{i\alpha}\) the Fresnel reflection coefficient of interface
\(i\). For polar materials supporting surface phonon-polaritons within
the Planck window, the dominant contribution generally originates from
the \(p\)-polarized evanescent modes.
The finite off-diagonal tensor (\ref{eq:currentcross}) generates
interference terms in the Poynting-vector average that are absent from
the diagonal independent-source sector of conventional FE.
For planar geometries, the Weyl representation of the Green tensors
reduces the problem to an integration over the parallel wavevector
$k_\parallel$.
\begin{figure}
\centering
\includegraphics[angle=0,scale=0.4]{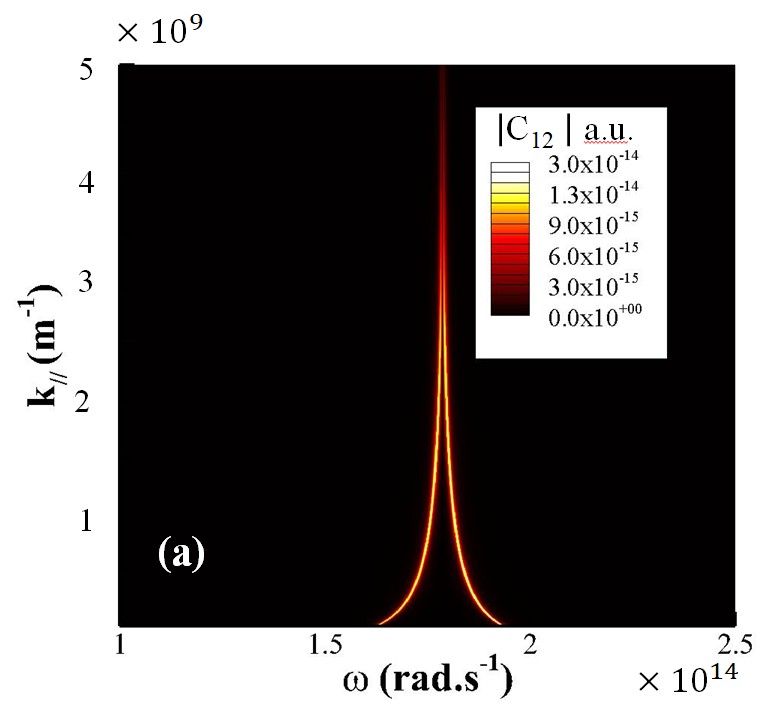}
\includegraphics[angle=0,scale=0.4]{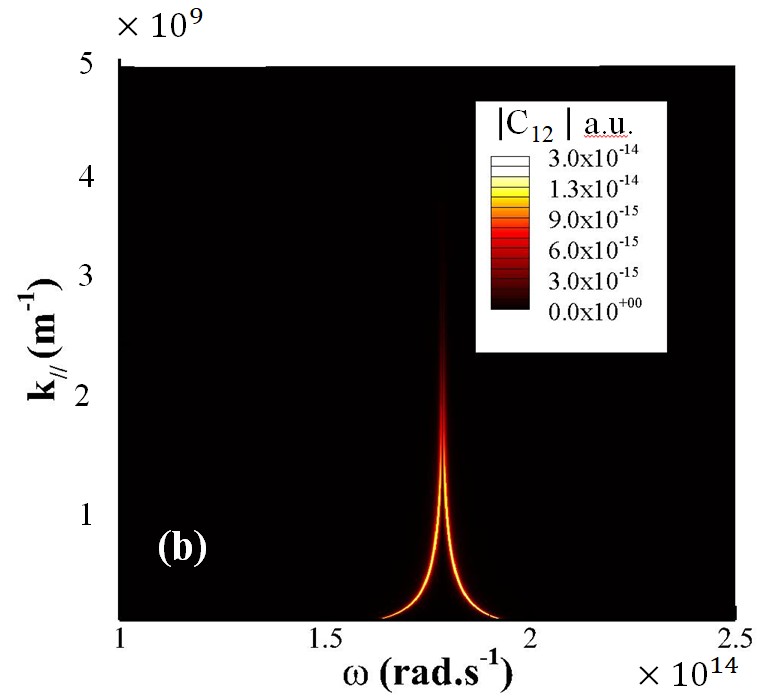}
\caption{Cross-correlation spectrum between the fluctuations of the two coupled surface oscillators for two SiC half-spaces of temperature $T_1=310\;K$ and $T_2=300\;K$ separated by a vacuum gap (a) $d=1\:nm$ and (b) $d=10\:nm$.
}
\end{figure}
The directional correlation-induced contribution follows directly from
Eq.~(\ref{eq:currentcross}); the measurable correction is obtained by
combining the reciprocal processes $1\rightarrow2$ and
$2\rightarrow1$ consistently with detailed balance.
Using Eq.~(\ref{eq:currentcross}), the directional contribution
$1\rightarrow 2$ reads~\cite{SupplMat}
\begin{equation}
\begin{aligned}
\Phi_{1\rightarrow2}^{\rm cross}(\omega)
=
-\frac{\omega^3\mu_0}{\pi}
\operatorname{Im}
\sum_{\alpha=s,p}
\int_0^\infty
k_\parallel
\,dk_\parallel
\,
\mathcal G^{(\alpha)}_{\beta\delta}
(\omega,k_\parallel,d)
\\
\times
\frac{
K(k_\parallel,d)
}{
|D_1D_2-K^2(k_\parallel,d)|^2
}
\\
\times
\Big[
(\mathbf Q_1\otimes\mathbf Q_2^\dagger)_{\beta\delta}
(D_2S_1+D_1^*S_2)
\\
+
(\mathbf Q_2\otimes\mathbf Q_1^\dagger)_{\beta\delta}
(D_1S_2+D_2^*S_1)
\Big],
\end{aligned}
\label{eq:explicit_cross_flux}
\end{equation}
where the quantity
\begin{equation}
\mathcal G^{(\alpha)}_{\beta\delta}
=
g^{E,(\alpha)}_{x\beta}
g^{H,(\alpha)*}_{y\delta}
-
g^{E,(\alpha)}_{y\beta}
g^{H,(\alpha)*}_{x\delta},
\end{equation}
represents the electromagnetic coupling factor in  polarization $\alpha$. The quantities $g^{E,H}$ which appear in this expression are the Weyl representations of the electric and magnetic Green tensors for the planar geometry, evaluated between the two interacting interfaces separated by the vacuum gap $d$. 

The net correlation-induced flux is
\begin{equation}
\varphi_{\rm cross}
=
\Phi_{1\rightarrow2}^{\rm cross}
-
\Phi_{2\rightarrow1}^{\rm cross},
\label{eq:phi_varphi_relation}
\end{equation}
where $\Phi_{2\rightarrow1}^{\rm cross}$ follows by exchanging
$1\leftrightarrow2$.

At equilibrium, finite cross correlations may persist, but reciprocity
and detailed balance impose
\begin{equation}
\varphi_{\rm FE}(\omega,T,T)
=
\varphi_{\rm cross}(\omega,T,T)
=
0.
\label{eq:equilibrium}
\end{equation}
Thus a nonzero covariance does not imply a net equilibrium heat flux.
The physically observable spectral heat flux is therefore
\begin{equation}
\varphi_{\mathrm{tot}}(\omega)
=
\varphi_{\mathrm{FE}}(\omega)
+
\varphi_{\mathrm{cross}}(\omega),
\label{eq:totalflux}
\end{equation}
which must remain positive for
$T_1>T_2$.
Although the conventional and correlation-induced contributions probe
the same hybridized surface modes, they arise from different statistical
contractions. The Green-tensor Fabry--P\'erot denominator describes the
resonant electromagnetic susceptibility, whereas the cross covariance
contains the additional mixing factor
\begin{equation}
\frac{K(k_\parallel,d)}
{|D_1D_2-K^2(k_\parallel,d)|^2}.
\label{eq:Fosc}
\end{equation}
The denominator determines the collective poles, while the numerator
$K$ describes mode mixing. Hence sharing the same hybridized poles does
not make the two statistical contributions identical; in particular,
$K\rightarrow0$ suppresses the off-diagonal covariance.

The interfacial coupling coefficient $K$ is determined directly from the
hybridized surface phonon-polariton dispersion
\begin{equation}
1-r_p^2(\omega)e^{-2k_\parallel d}=0,
\end{equation}
where
\begin{equation}
r_p(\omega)
=
\frac{\epsilon(\omega)-1}
{\epsilon(\omega)+1}
\end{equation}
is the Fresnel reflection coefficient for $p$ polarization in the
electrostatic limit. The corresponding symmetric and antisymmetric
surface-mode frequencies are denoted
$\Omega_+(k_\parallel,d)$ and
$\Omega_-(k_\parallel,d)$, respectively~\cite{SupplMat}.
The coupling strength is then obtained from the splitting of these
hybridized modes,
\begin{equation}
K(k_\parallel,d)
=
\frac{M}{2}
\left[
\Omega_+^2(k_\parallel,d)
-
\Omega_-^2(k_\parallel,d)
\right].
\label{eq:Kdispersion}
\end{equation}
Equation~(\ref{eq:Kdispersion}) identifies the deterministic coupling
from the independently calculated Maxwell spectrum; no Langevin
statistics enters this determination. Thus
$\Omega_\pm\rightarrow K\rightarrow\langle X_1X_2^*\rangle$, and the
cross correlation is not assumed in determining $K$.
Equation~(\ref{eq:Kdispersion}) provides a direct connection between the
electromagnetic description of the coupled interfaces and the
microscopic oscillator model. Larger mode splittings correspond to
stronger interfacial coupling and, consequently, to larger
fluctuating-current cross correlations.
Figure~1(a) shows the dispersion of the hybridized surface
phonon-polariton modes supported by two SiC half-spaces separated by a
vacuum gap. As the separation distance decreases, the splitting between
the symmetric (\(\Omega_+\)) and antisymmetric (\(\Omega_-\)) branches
increases, reflecting the growing electromagnetic coupling between the
two interfaces. 
The corresponding interfacial coupling coefficient
$K(k_{\parallel},d)$ is displayed in Fig.~1(b). Consistent with the
increasing mode splitting, $K$ grows rapidly as the gap narrows and
reaches its largest values in the extreme near-field regime. Moreover,
the coupling remains significant over an increasingly broad range of
wavevectors as the separation decreases, demonstrating the emergence of
collective surface excitations extending coherently across the vacuum
gap.
The resulting oscillator cross-correlation spectrum is shown
in Fig.~2. A pronounced resonance appears near the surface
phonon-polariton frequency, where the hybridized modes are
most strongly excited. The amplitude of the cross-correlation
increases markedly as the separation distance decreases from
$10\;nm$ to $1\;nm$, reflecting the growth of the interfacial
coupling coefficient $K$. These results demonstrate that substantial cross correlations emerge
despite the statistical independence of the Langevin forces. Their
growth with decreasing separation directly reflects the progressive
delocalization of the surface excitations and the formation of
collective modes extending across the vacuum gap.
\begin{figure}
\centering
\includegraphics[angle=0,scale=0.4]{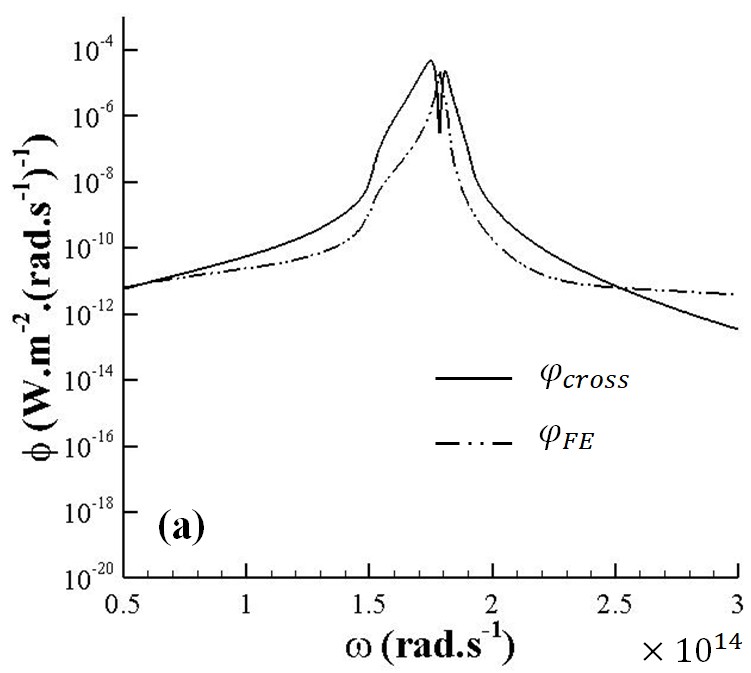}
\includegraphics[angle=0,scale=0.4]{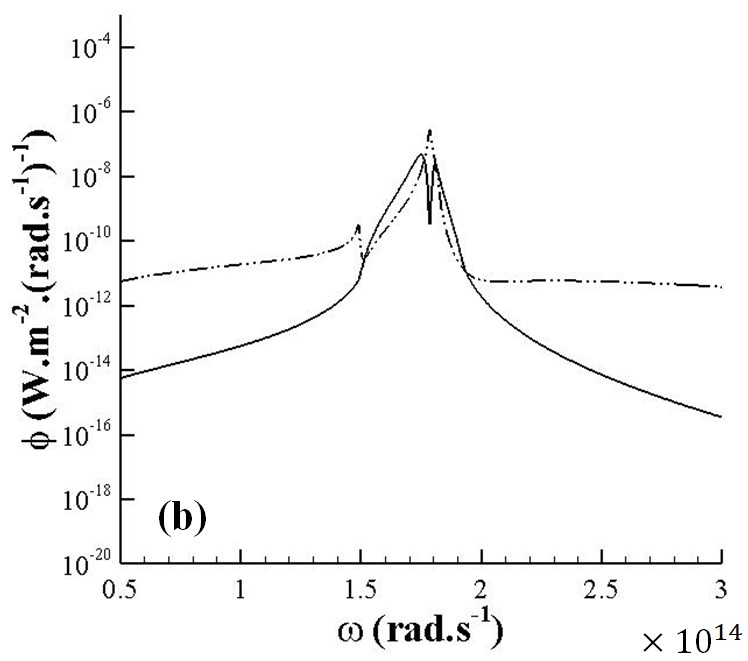}
\includegraphics[angle=0,scale=0.4]{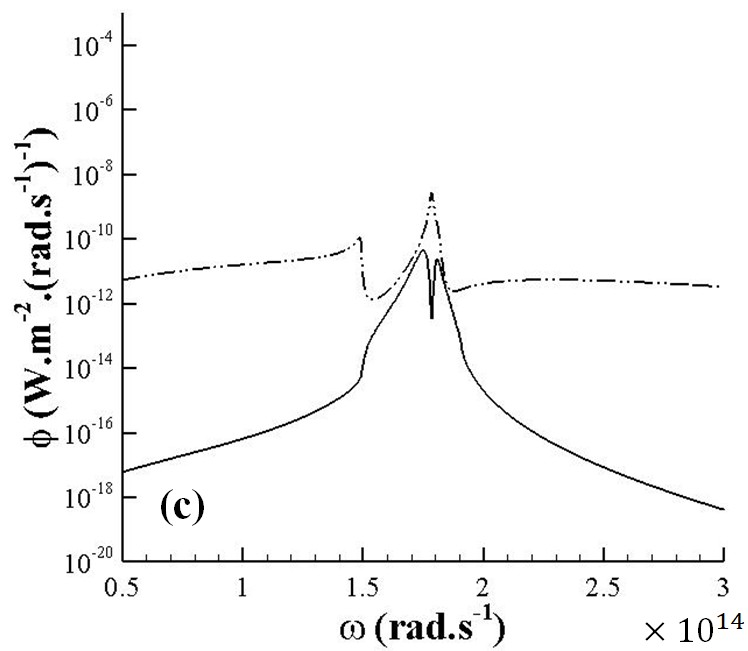}
\caption{Spectral radiative heat flux carried by correlated fluctuations (full line) and predicted by conventional FE (dashed line) between two SiC half-spaces separated by a vacuum gap (a) $d=1\;nm$, (b) $d=10\;nm$ and (c) $d=100\;nm$.}
\end{figure}

To quantify the importance of the correlation-induced contribution on the heat transfer, we
introduce the spectral ratio
\begin{equation}
R(\omega,d)
=
\frac{
\varphi_{\mathrm{tot}}
}{
\varphi_{\mathrm{FE}}
}
=
1+
\frac{
\varphi_{\mathrm{cross}}
}{
\varphi_{\mathrm{FE}}
},
\label{eq:Rdef}
\end{equation}
where both fluxes are integrated over the parallel wavevector
$k_\parallel$. Since
$\varphi_{\mathrm{cross}}$
originates from an interference term, it is not necessarily positive. 
However, according to the second principle, the total flux must remain positive when
$T_1>T_2$.

The spectral heat fluxes predicted by conventional
FE and by the correlation-induced
contribution are compared in Fig.~3.
For nanometric and subnanometric gaps, the correlation-induced
contribution becomes comparable to the conventional FE flux near the
surface-phonon-polariton resonance. This behavior reflects the onset of
collective fluctuations associated with the hybridized surface modes.
As the separation increases, the mode hybridization weakens and the
cross-correlation contribution rapidly disappears, recovering the
independent-source limit assumed in conventional FE.

\begin{figure}
\centering
\includegraphics[angle=0,scale=0.4]{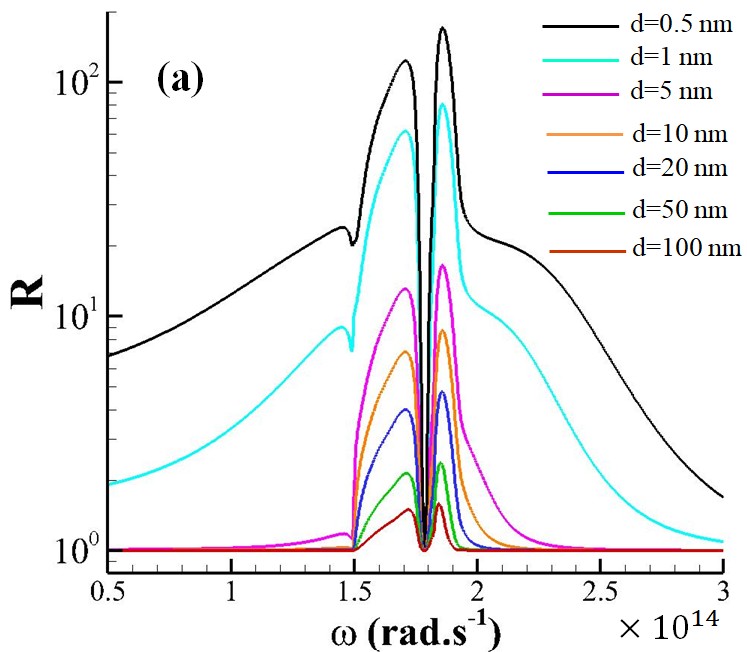}
\includegraphics[angle=0,scale=0.4]{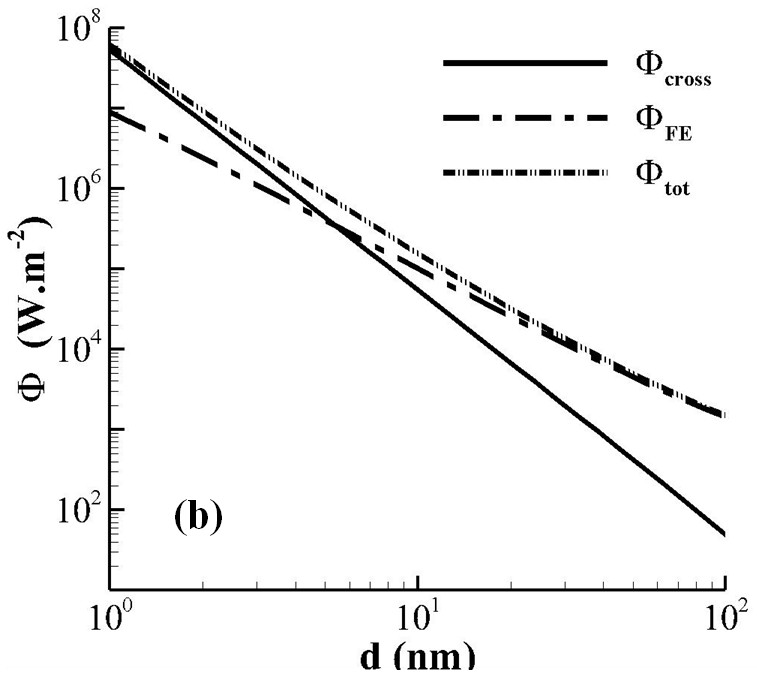}
\caption{
(a) Spectral amplification factor
$R(\omega,d)=\varphi_{\rm tot}(\omega,d)/
\varphi_{\rm FE}(\omega,d)$
for two SiC half-spaces at $T_1=310\,\mathrm{K}$
and $T_2=300\,\mathrm{K}$, as a function of angular
frequency $\omega$ for various separation distances $d$.
(b) Frequency-integrated heat fluxes $\Phi_{\rm cross}$, $\Phi_{\rm FE}$ and  $\Phi_{\rm tot}$ as functions of the vacuum-gap width $d$.
}
\end{figure}

For nearly identical modes,
$g\simeq |K|/(2M\Omega_0)$ and
$\Delta\Omega=|\Omega_+-\Omega_-|\simeq2g$. The strongly hybridized
regime is reached when $\Delta\Omega\gtrsim\gamma$, where the
correlation-induced contribution becomes significant.

Figure~4(a) illustrates the correlation-induced amplification
of radiative heat transfer. A pronounced enhancement occurs
within the Reststrahlen band, where surface phonon-polaritons
mediate strong interfacial coupling. As the gap narrows,
the amplification increases rapidly, exceeding one order
of magnitude at subnanometric separations. Remarkably,
the largest enhancement factors occur slightly away from
the surface phonon-polariton resonance, where the conventional
FE contribution decreases more rapidly than the
correlation-induced correction. This spectral behavior
provides a characteristic signature of correlated surface
fluctuations beyond the independent-source approximation.

To assess the impact of these correlations on the measurable
heat exchange, we consider the frequency-integrated flux
\begin{equation}
\Phi_{\rm tot}(d)
=
\int_0^\infty\frac{d\omega}{2\pi}
\left[
\varphi_{\rm FE}(\omega,d)
+
\varphi_{\rm cross}(\omega,d)
\right].
\label{eq:integratedflux}
\end{equation}
As shown in Fig.~4(b), the conventional FE contribution
follows the characteristic $\Phi_{\rm FE}\propto d^{-2}$
scaling, whereas the total flux exhibits a steeper
$\Phi_{\rm tot}\propto d^{-3}$ dependence below approximately
$10\,\mathrm{nm}$, where cross-correlated fluctuations
dominate the energy exchange. At larger separations,
the interfacial coupling weakens, the correlation-induced
contribution becomes negligible, and the conventional
$d^{-2}$ scaling is recovered. This crossover identifies
a distinct extreme near-field transport regime and provides
a potentially more discriminating signature of correlated
thermal fluctuations than the flux enhancement alone.

Experimentally, this prediction could be tested by measuring
the distance dependence of the heat exchange and comparing
it with conventional FE. A particularly stringent test
would combine heat-transfer measurements with spectroscopy
of the symmetric and antisymmetric surface-polariton modes.
Their frequency splitting would independently determine
the coupling coefficient $K(k_\parallel,d)$, allowing
the correlation-induced heat flux to be predicted without
fitting the heat-transfer data. At subnanometric separations,
however, additional transport mechanisms, particularly
phonon tunneling~\cite{GomezViloria,Guo2023}, may coexist
with radiative exchange. Distinguishing these contributions
therefore requires examining their spectral signatures
and distance-scaling behavior rather than relying solely
on the magnitude of the measured heat flux.

To conclude, we have identified a correlation-induced contribution to
extreme near-field radiative heat transfer arising from an off-diagonal
covariance of interacting surface fluctuations, a statistical sector
absent from the conventional independent-source Rytov description.
The effect becomes significant when the surface-mode splitting
approaches the intrinsic linewidth and leads to experimentally testable
deviations from conventional FE. While electromagnetic propagation
remains described by the full Green tensor, the proposed extension
concerns the covariance structure of the effective fluctuating surface
degrees of freedom. These results provide a framework for identifying
correlated thermal transport in strongly coupled nanoscale systems.

\begin{acknowledgments}
\end{acknowledgments}

\end{document}